\documentclass[11pt]{article}

\usepackage{amsmath}
\usepackage{graphicx}
\usepackage{amsfonts}
\usepackage{amssymb}
\usepackage{epsfig}
\usepackage{color}
\usepackage{psfrag}
\usepackage{epstopdf}

\newcommand{\EQ}{\begin{equation}}
\newcommand{\EN}{\end{equation}}
\newcommand{\be}{\begin{equation}}
\newcommand{\ee}{\end{equation}}
\newcommand{\bea}{\begin{eqnarray}}
\newcommand{\eea}{\end{eqnarray}}

\begin{document} \setcounter{page}{0}
\topmargin 0pt
\oddsidemargin 5mm
\renewcommand{\thefootnote}{\arabic{footnote}}
\newpage
\setcounter{page}{0}
\topmargin 0pt
\oddsidemargin 5mm
\renewcommand{\thefootnote}{\arabic{footnote}}
\newpage
\begin{titlepage}
\vspace{0.5cm}
\begin{center}
{\large {\bf Correlation spreading and properties of the quantum state}\\
{\bf in quench dynamics }}\\
\vspace{1.8cm}
{\large Gesualdo Delfino}\\
\vspace{0.5cm}
{\em SISSA -- Via Bonomea 265, 34136 Trieste, Italy}\\
{\em INFN sezione di Trieste}\\
\end{center}
\vspace{1.2cm}

\renewcommand{\thefootnote}{\arabic{footnote}}
\setcounter{footnote}{0}

\begin{abstract}
\noindent
The light cone spreading of correlations following a quantum quench is obtained from first principles. Fully taking into account quantum and interaction effects, the derivation shows how light cone dynamics does not require peculiar properties of the post-quench state.
\end{abstract}
\end{titlepage}

\section{Introduction}
Information does not propagate instantaneously. Even for extended quantum systems in which relativistic effects can be ignored, so that the finite speed of light does not enter the theoretical treatment, upper bounds on the speed of signal propagation can be obtained \cite{LR}. As a consequence,  correlations grow significantly only within ``light" cones in space-time. In recent years substantial attention has been devoted to correlation spreading in isolated extended quantum systems brought out of equilibrium. A main reason is that the quantum state of such systems (see e.g. \cite{PSSV,EFG,dAKPR,CEM} for reviews) proved difficult to characterize, so that identifying the mechanism of correlation spreading can possibly help elucidating its properties. The following picture was proposed in \cite{CC} (see also \cite{CC_review}). Following a sudden alteration of the Hamiltonian (``quantum quench") bringing the system out of equilibrium, pairs of particle excitations with opposite momenta are created and travel classically with a maximal velocity $v_m$ and without scattering; then correlations between two points separated by a distance $r$ start to develop at time $t\simeq r/2v_m$, when they are first and simultaneously reached by two particle excitations emitted at the same point. The assumptions associated to the picture aim at accounting for the light cone effect exhibited by the examples studied analytically in \cite{CC}. A problematic point, from the point of view of the general interpretation, is that such examples are also those for which the strong assumption of propagation without scattering effectively holds. Indeed, the analytic results were obtained for two different cases. The first corresponds to mass quenches in a free theory, which indeed produce pairs of non-interacting particles with opposite momenta (see below). The second case is that of integrable (in particular conformal) dynamics on a half plane space-time, where the state is again made of pairs of particles with opposite momenta; here the particles can interact, but integrability ensures that the scattering preserves number of particles and individual momenta \cite{GZ}. Notice that, within the above picture, integrability in one spatial dimension appears as a necessary condition, since particles moving classically on a line cannot miss each other. Also the experimental observation of the light cone within a one-dimensional system was described in terms of a post-quench state with excitations organized in pairs \cite{Cheneau}. This state of affairs leaves open two questions that we address in this paper. First: is the light cone effect related to specific features of the quantum state? Second: can the light cone be derived from first principles?

\section{The post-quench state}
It is clear from the above discussion that a crucial issue is that of having theoretical access to a larger class of quenches. It was shown in \cite{quench} and further illustrated in \cite{quench2} that a fundamental and general approach can be formulated {\it near} quantum critical points. This condition ensures that the correlation length is sufficiently large (both before and after the quench), so that the problem is described by massive quantum field theory. We now briefly recall the formalism and the result for the post-quench state. Before the quench the translationally invariant system, with Hamiltonian $H_0$, is in the ground state, which is the vacuum state $|0\rangle$ of the pre-quench field theory. Excitations over this vacuum state correspond to particles with momentum ${\bf p}$ and energy $E_{\bf p}=\sqrt{{\bf p}^2+M^2}$, $M$ being the particle mass. The post-quench Hamiltonian can be written as
\EQ
H=H_0+\lambda\int d{\bf x}\,\Psi\,,
\label{H}
\EN
where we refer to $\lambda$ and $\Psi$ as the quench parameter and the quench operator, respectively. Due to the quench at $t=0$, the system passes in the new state $|\psi_0\rangle$, which is determined in the scattering formalism and, to first order in $\lambda$, reads \cite{quench,quench2}
\bea
|\psi_0\rangle &\simeq & \left(1-i\lambda\int_0^\infty dt\int_{-\infty}^\infty d{\bf x}\,\Psi({\bf x},t)\right) |0\rangle \nonumber\\
&=& |0\rangle+\lambda\sum_{n=1}^\infty\frac{(2\pi)^D}{n!}\int_{-\infty}^{\infty}\prod_{i=1}^n d{\bf p}_i\,\delta(\sum_{i=1}^n{\bf p}_i)\,\frac{[F_n^\Psi({\bf p}_1,\ldots,{\bf p}_n)]^*}{\sum_{i=1}^nE_{{\bf p}_i}}\,|{\bf p}_1,\ldots,{\bf p}_n\rangle\,,
\label{psi0}
\eea
where $D$ is the number of spatial dimensions and we used the relation 
\EQ
\Psi({\bf x},t)=e^{i{\bf P}\cdot{\bf x}+iH_0t}\Psi(0,0)e^{-i{\bf P}\cdot{\bf x}-iH_0t}\,,
\label{shift}
\EN
${\bf P}$ being the momentum operator. The result has been expanded over the complete basis of multiparticle states\footnote{With respect to \cite{quench,quench2}, we lighten the notation adopting a different normalization of states. We also recall that quantum field theory and its particle formalism automatically implement properties such as locality of interactions and cluster decomposition of correlators (see e.g. \cite{Weinberg}).} $|{\bf p}_1,\ldots,{\bf p}_n\rangle$ of the pre-quench theory, introducing the matrix elements (form factors)
\EQ
F_n^\Psi({\bf p}_1,\ldots,{\bf p}_n)=\langle 0|\Psi(0,0)|{\bf p}_1,\ldots,{\bf p}_n\rangle;
\label{ff}
\EN
the states entering the scattering formalism are {\it asymptotic} states with particles far apart from each other and eigenvalues of energy and momentum given by $\sum_iE_{{\bf p}_i}$ and $\sum_i{\bf p}_i$, respectively.

 Notice that if both $H_0$ and $\Psi$ are quadratic operators, as it is the case for a mass quench for free particle excitations, $F_n^\Psi\propto\delta_{n,2}$ and only the contribution of $|{\bf p},-{\bf p}\rangle$ survives in (\ref{psi0}) (more pairs are generated at higher orders in $\lambda$). In the other cases, i.e. in presence of interacting particle excitations, the form of $|\psi_0\rangle$ consisting of particle pairs with opposite momenta does {\it not} occur.

\section{Derivation of the light cone}
In principle, the equal-time two-point function $\langle\psi_0|\Phi({\bf x},t)\Phi(0,t)|\psi_0\rangle$  for an operator $\Phi$, which contains the information about the spreading of correlations over a distance $|{\bf x}|$ after a time $t$ from the quench, can be calculated to first order in $\lambda$ using the expression (\ref{psi0}), and to higher orders continuing the perturbative expansion. In this paper, however, we are not interested in an explicit calculation of two-point functions, which is essentially out of reach for the general case we are addressing. Instead, we are interested in a specific property, the light cone, and will show that this can be derived non-perturbatively, relying only on first principles. For this purpose, we retain from the perturbative result (\ref{psi0}) only the fact that in general the post-quench state is a superposition over all multiparticle states, with the only constraint that their {\it total} momentum is zero. Hence, we perform the derivation for the {\it general state} 
\EQ
|A\rangle=\sum_{n=0}^\infty\int_{-\infty}^{\infty}\prod_{i=1}^n d{\bf p}_i\,\delta(\sum_{i=1}^n{\bf p}_i)\,f_n({\bf p}_1,\ldots,{\bf p}_n)\,|{\bf p}_1,\ldots,{\bf p}_n\rangle\,,
\label{A}
\EN
which features generic coefficients $f_n$ for the superposition\footnote{Notice that, since $|A\rangle$ is a non-perturbative state of the post-quench theory, the superposition is taken over the basis of asymptotic states of the post-quench excitations. In (\ref{psi0}), instead, the states entering the superposition are those of the pre-quench theory, and the post-quench mass is reconstructed order by order in perturbation theory.}. We then obtain
\bea
\langle A|\Phi({\bf x},t)\Phi(0,t)|A\rangle &=& \sum_{n_1,n_2,m=0}^\infty\int \prod_{i=1}^{n_1}d{\bf p}_i\prod_{j=1}^{n_2}d{\bf p}'_j \prod_{k=1}^{m}d{\bf q}_k\,f^*_{n_2}({\bf p}'_1,\ldots,{\bf p}'_{n_2})\,f_{n_1}({\bf p}_1,\ldots,{\bf p}_{n_1})\, \nonumber\\
&\times& F_{n_2,m}^\Phi({\bf p}'_1,\ldots,{\bf p}'_{n_2}|{\bf q}_1,\ldots,{\bf q}_m)\,F_{m,n_1}^\Phi({\bf q}_1,\ldots,{\bf q}_m|{\bf p}_1,\ldots,{\bf p}_{n_1})\nonumber\\
&\times& \delta(\sum_{i=1}^{n_1}{\bf p}_i)\,\delta(\sum_{j=1}^{n_2}{\bf p}'_j)\,e^{-i\varphi({\bf x},t)}\,,
\label{expansion}
\eea
where we inserted a complete set of asymptotic $m$-particle states in between the two operators, introduced the notation
\EQ
F_{m,n}^\Phi({\bf q}_1,\ldots,{\bf q}_m|{\bf p}_1,\ldots,{\bf p}_n)=\langle {\bf q}_1,\ldots,{\bf q}_m|\Phi(0,0)|{\bf p}_1,\ldots,{\bf p}_n\rangle\,,
\label{fmn}
\EN
and use (\ref{shift}) to obtain
\EQ
\varphi({\bf x},t)={\bf x}\cdot\sum_{k=1}^{m}{\bf q}_k+t\left(\sum_{i=1}^{n_1}E_{{\bf p}_i}-\sum_{j=1}^{n_2}E_{{\bf p}'_j}\right)\,.
\label{phase}
\EN

\begin{figure}[t]
\begin{center}
\includegraphics[width=10cm]{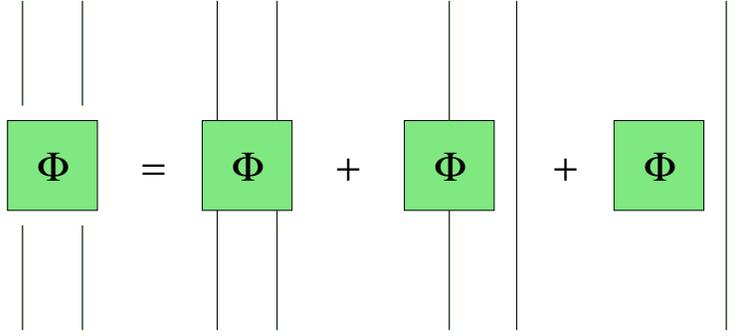}
\caption{Connectedness structure of the matrix element $F_{2,2}^\Phi$.}
\label{connectedness}
\end{center}
\end{figure}

The term with $m=0$ in the expansion (\ref{expansion}) is ${\bf x}$-independent, hence contributes to the disconnected part of the two-point function and can be ignored in the discussion of spatial correlations. For $m>0$, let us consider the limit of large $|{\bf x}|$ ($|{\bf x}|\gg 1/M$). Then the phase in (\ref{expansion}) rapidly oscillates and suppresses the integral, unless there is a stationary point, i.e. $\nabla_{{\bf q}_k}\varphi=0$ for $k=1,2,\ldots,m$. A superficial examination of (\ref{expansion}) and (\ref{phase}) may suggest that stationarity is not satisfied, with the consequence that no sizeable correlation arises between the two points at any time. However, we have to remember that the matrix elements (\ref{fmn}) actually decompose into the sum of a fully connected part plus disconnected contributions \cite{Weinberg}, namely
\bea
F_{m,n}^\Phi({\bf q}_1,\ldots,{\bf q}_m|{\bf p}_1,\ldots,{\bf p}_n) &=& \langle {\bf q}_1,\ldots,{\bf q}_m|\Phi(0,0)|{\bf p}_1,\ldots,{\bf p}_n\rangle_\textrm{connected}
\label{disc}\\
&+& \delta({\bf q}_1-{\bf p}_1)\langle {\bf q}_2,\ldots,{\bf q}_m|\Phi(0,0)|{\bf p}_2,\ldots,{\bf p}_n\rangle_\textrm{connected}+\cdots\,,\nonumber
\eea
where the dots stay for all remaining contractions of particles from the state on the left with particles from the state on the right (see figure~\ref{connectedness}). The relevant point is that each contraction yields a delta function $\delta({\bf q}_k-{\bf p}_i)$. It follows that for $m=1$ the disconnected contribution in the second line of (\ref{disc}) modifies (\ref{phase}) into
\EQ
{\bf x}\cdot{\bf q}+t\left(E_{\bf q}+\sum_{i=2}^{n_1-1}E_{{\bf p}_i}+E_{-{\bf q}-({\bf p}_2+\cdots+{\bf p}_{n_1-1})}-\sum_{j=1}^{n_2}E_{{\bf p}'_j}\right)\,,
\label{phase2}
\EN
where we made explicit that the constraint $\sum_{i=1}^{n_1}{\bf p}_i=0$ induces the presence of ${\bf q}$ in two energy terms. Differentiating now with respect to ${\bf q}$ one obtains the stationarity condition
\EQ
{\bf x}=-{{\bf V}}t\,,
\label{stationarity}
\EN
where 
\bea
{\bf V} &=& {\bf v}_{\bf q}+{\bf v}_{\bf q+p_2+\cdots+p_{n-1}}\,,\\
{\bf v}_{\bf p} &\equiv & \nabla_{\bf p}E_{\bf p}=\frac{\bf p}{\sqrt{{\bf p}^2+M^2}}\,.
\eea
In our natural units the maximal value of the velocity $|{\bf v}_{\bf p}|$ is one, so that upon integration over momenta the values of $|{\bf V}|$ span the interval $(0,2)$. It follows that the stationarity condition (\ref{stationarity}) is satisfied when
\EQ
t>\frac{|{\bf x}|}{2}\,.
\label{cone}
\EN
This conclusion was obtained considering the intermediate state with $m=1$. For $m>1$ a stationary phase is provided by disconnected contributions yielding $m$ delta functions. The stationarity condition with respect to each momentum ${\bf q}_k$ is again (\ref{stationarity}) with a maximal value of $|{\bf V}|$ which remains 2; as a consequence (\ref{cone}) is not modified. For $|{\bf x}|$ large enough the contribution to the connected correlator from configurations not producing a stationary point vanishes. Since the condition for stationarity is the same for all the terms allowing for it, for $|{\bf x}|$ large enough spatial correlations exist only within the light cone specified by (14).

\section{Discussion}
We see that the light cone (\ref{cone}) is a general property of two-point functions (\ref{expansion}) over states of the form (\ref{A}) with $f_n\neq 0$ for some $n>1$, a condition required for having suitable disconnected contributions. The factor $1/2$ in (\ref{cone}) is related to the zero momentum condition on the state $|A\rangle$, corresponding to spatial translation invariance. The fact that the state (\ref{psi0}) is only a particular case of $|A\rangle$ leaves room for the same effect to be observed within paths to non-equilibrium different from an instantaneous alteration of the Hamiltonian. 

It is worth stressing how our general derivation of the light cone relied only on first principles of quantum field theory and involved no approximations. Of course, being general, the result coincides with that obtained from explicit calculations of two-point functions in solvable cases (see \cite{CC_review}). In particular, the result that the light cone is a property of the individual terms of the series (\ref{expansion}) and does not require its resummation generalizes that observed in \cite{CEF} for the order parameter correlator in the transverse field Ising chain, a case in which the particles do not interact; obviously, resummation would be needed to obtain the functional form of the connected correlator inside the light cone. 

The derivation makes clear the essential role played for the light cone result (\ref{cone}) by the delta functions over momenta explicitly appearing in (\ref{A}) and (\ref{disc}). It is worth stressing that these delta functions follow from first principles (momentum conservation and connectedness structure), and that no other delta function can appear in the general interacting case we consider. As we already stressed, only if the theory is free both before and after the quench the post-quench state will be made of pairs of particles with opposite momenta, a structure corresponding to delta functions $\delta({\bf p}_i+{\bf p}_j)$ in the coefficients $f_n$ entering (\ref{A}); in (\ref{expansion}) these give rise to some terms containing squares of delta functions that need to be regularized. The regularization (see \cite{CEF}) shows that the location of the light cone in this particular case coincides with that we are now deriving for the interacting case. Similarly, we see that the result (\ref{cone}) does not depend on the detailed manipulation of the annihilation poles of form factors required for the explicit calculations of correlators in solvable cases (see e.g. \cite{CEF}): it is well known \cite{Smirnov} that those poles are in one to one correspondence with the delta functions in (\ref{disc}), and that they cannot give rise to any extra delta function able to affect the light cone. 

We see that the light cone is in no way related to states with a specific structure of excitations or a peculiar propagation mode. Equation (\ref{psi0}) shows that in generic dimension the organization in pairs arises only when the particle excitations do not interact. Interaction, on the other hand, poses no difficulty once the quantum nature of the problem is taken into account: it is the non-trivial connectedness structure of the matrix elements which produces the final result, without assumptions on scattering properties. 

This also makes theoretically clear that in one dimension the light cone is in no way related to integrability, which in the heuristic picture seems necessary to reconcile diffusionless propagation with interaction on the line. Actually, it was already shown in \cite{quench} that the non-equilibrium setting substantially reduces the room for integrability. Indeed, in the field theory describing a quench, which in general allows for transmission of energy (and then of particle excitations) from pre- to post-quench times, factorization of scattering amplitudes (and integrability with it) is not compatible with particle interaction \cite{quench} (see also \cite{Schuricht}). In other words, the non-trivial conserved currents associated with integrability at equilibrium do not survive the quench if the particle excitations interact\footnote{Exceptions cannot be excluded on the lattice if the pre- and/or post-quench states do not admit a continuum limit due to a small correlation length. The argument based on factorization of scattering amplitudes also leaves room for exceptions if the particle excitations near criticality do not admit a relativistic dispersion relation.}. Integrability in presence of interaction requires eliminating transmission, namely giving up the notion of a pre-quench Hamiltonian $H_0$ and going back to the half plane space-time of \cite{GZ,Cardy_boundary_CFT}. A recent survey \cite{PPV} of results for spin chains confirms the conclusions of \cite{quench} on integrability. 

It is relevant that the deviation of the post-quench state from the structure with excitations organized in pairs has measurable implications. It was shown in \cite{quench} that in one dimension the state (\ref{psi0}) with $F_1^\Psi\neq~0$ allows for undamped oscillations of one-point functions\footnote{Undamped oscillations were analytically derived in \cite{quench} for small quenches, showing also that they certainly persist up to a time scale that goes to infinity as the size of the quench is reduced. This has to be contrasted, in particular, with the case of post-quench states with excitations organized in pairs, for which oscillations are damped already for small quenches \cite{quench,quench2}.}, a feature which does not easily fit within the usual expectations about relaxation in isolated one-dimensional systems (see \cite{CEM}). It was pointed out in \cite{quench} that the simplest realization of this phenomenon arises when suddenly switching on a small longitudinal field starting from the paramagnetic phase of the Ising spin chain. The predicted undamped oscillations of the order parameter have been numerically observed in \cite{RMCKT,KCTC}. The agreement between theory and numerical data is further illustrated in \cite{quench2}. 

In summary, the light cone spreading of correlations in quantum quenches near criticality has been generally derived from first principles. The derivation fully incorporates quantum and interaction effects, and disentangles light cone dynamics from assumptions on the properties of the post-quench state.

\end{document}